\documentclass[10pt,pre,twocolumn,aps,superscriptaddress]{revtex4-1}
\usepackage{epsfig,amsmath,amssymb,graphicx,color,calc,epstopdf}

\let\e=\varepsilon   
\let\l=\lambda    
\let\s=\sigma

\def\to{\rightarrow}

\newcommand{\beq}{\begin{equation}}
\newcommand{\eeq}{\end{equation}}
\newcommand{\ba}{\begin{align}}
\newcommand{\ea}{\end{align}}

\def\de{\mathrm d}

\begin{document}

\title{The Gardner transition in finite dimensions}

\author{Pierfrancesco Urbani}
\affiliation{
IPhT, CEA/DSM-CNRS/URA 2306, CEA Saclay, F-91191 Gif-sur-Yvette Cedex, France 
}

\author{Giulio Biroli}
\affiliation{
IPhT, CEA/DSM-CNRS/URA 2306, CEA Saclay, F-91191 Gif-sur-Yvette Cedex, France 
}

\begin{abstract}
Recent works on hard spheres in the limit of infinite dimensions revealed that glass states, envisioned as meta-basins in configuration space, can break up in a multitude of separate basins at low enough temperature or high enough pressure, leading to the emergence of new kinds of soft-modes and unusual properties.
In this paper we study by perturbative renormalisation group techniques the critical properties of this transition, which has been discovered in disordered mean-field models in the '80s. We find that the upper critical dimension $d_u$ above which mean-field results hold 
is strictly larger than six and apparently non-universal, i.e. system dependent. Below $d_u$, we do not find any perturbative attractive fixed point (except for a tiny region of the 1RSB breaking parameter), thus showing that the transition in three dimensions either is governed by a non-perturbative fixed point unrelated to the Gaussian mean-field one or becomes first order or does not exist. We also discuss possible relationships with the behavior of spin glasses in a field.      
\end{abstract}


\maketitle

\begin{figure}
\scalebox{0.7}{\input{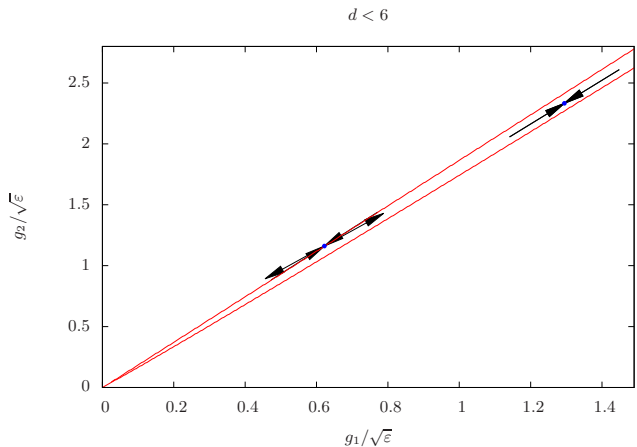}}
\caption{The RG fixed points in the quadrant $g_1>0$ and $g_2>0$ for $d<6$ and $m\in[m^*,\tilde m]$. We find an attractive fixed point and a partially repulsive fixed point on the border of its basin of attraction (continuous red line). For both fixed points, the eigenvectors directions of the linearized RG flow are almost collinear. For $m\in [\tilde m, 1]$ only the unstable fixed point survives.
}
\label{fig:dminore6}
\end{figure}

The properties of glasses at low temperatures are the subject of extensive experimental, 
numerical and analytical investigations. In order to understand them, one has to study  
the properties of the amorphous solids in which liquids freeze at the glass transition. 
Hence a crucial preliminary step is arguably understanding glass-formation.  
One of the most prominent theoretical approaches to do that is the Random First Order Transition (RFOT) theory introduced by Kirkpatrick, Thirumalai, and Wolynes \cite{KW87, KT87, KTW89, BB09}. It has its roots in the mean-field theory of disordered models but, as it has become
clear in recent years, it goes well beyond that. RFOT theory applies to all systems characterized by a certain kind of 
(free-)energy landscape, such that below a given temperature $T_d$ an exponential number (in the system size) of 
metastable states emerges. By lowering the temperature their thermodynamics becomes ruled by the competition between  
two kinds of contributions: one (free-energetic) that favours states with lower internal free energy because their corresponding Boltzmann weight is larger, and the other (entropic) which favours  states having high internal free energy because they are more numerous.
At the so called Kauzmann temperature, $T_K$, the entropic contribution vanishes and the system freezes 
in one low-lying glass state. RFOT theory advocates that this is precisely what happens for super-cooled liquids
approaching the glass transition, where $T_d$ corresponds to the so-called Mode Coupling cross-over and 
$T_K$ to the ideal glass transition. A major result of the last thirty years  
was to show that this is indeed the case within mean-field theory \cite{BB11}. Actually, the range of systems displaying such an energy landscape---at the mean field level---is remarkably broad: it encompasses physical systems such as super-cooled liquids, colloids, proteins \cite{WL12, KT14} and models central in other fields like random K-satisfiability \cite{KrzakalaBook}.
Whether this remains true beyond the mean-field approximation it is still a matter of debate, although there are by now 
remarkable numerical and experimental evidences \cite{BK13, WL12}. 
\\
For a long time the properties of low-temperature glasses remained a separate research subject from the much more studied problem of glass transition with the notable exception of Ref. \cite{LW07}. Recently, however,
there has been an increasing research effort aimed at understanding amorphous solids' unusual features and their relationship with the glass transition \cite{Bi14}. This was to great extent motivated by the study of jamming \cite{LN98, LNSW10}. In this context, a new twist of RFOT theory is the suggestion that glass states, envisioned as meta-basins in configuration space, can break up in a multitude of separate basins at low enough temperature or high enough pressure, leading to the emergence of new kinds of soft modes and unusual properties \cite{CKPUZ14}. This transition, called Gardner transition \cite{GKS85, Ga85}, was actually found long-time ago for several mean-field models characterized by a RFOT.
In these systems at a temperature, $T_K$, there is a glass phase transition at which, technically, 
a one replica symmetry breaking (1RSB) phase emerges and at a lower temperature, $T_G$, there is a Gardner 
transition towards a full replica symmetry breaking (FRSB) phase, see for example the case of the Ising $p$-spin disordered models \cite{GM84, MPV87}.
As pointed out in \cite{BFP97, MR03} (see also \cite{Ri13}) this transition from a valley in configuration space to a multitude of separated basins takes place also for non-equilibrium glass states. In consequence, it 
is not limited to the (unreachable) equilibrium regime below $T_K$ but is also relevant for common 
non-equilibrium protocols such as quenches or crunches during which the system gets trapped in a metastable state.   
For a very long time the study of the Gardner transition remained bounded to abstract mean-field models. From the point of view of the physics of glasses, it was just a pure intellectual curiosity. Recent works on the solution of glassy hard spheres in infinite dimensions highlighted its relevance for amorphous materials \cite{KPZ12, KPUZ13, jstatCKPUZ13, CKPUZ14}. 
In fact the FRSB phase appearing below $T_G$ is marginally stable and its soft-modes are deeply related to the unusual features displayed by jammed packings \cite{LNSW10, Wy12, WNW05, WSNW05, LDW13, BW06, BW07, BW09b}. Remarkably, the FRSB mean-field theory predicts values 
for the critical exponents of the jamming transition that are in perfect agreement with the ones observed in numerical
simulations in two and three dimensions.
In consequence, understanding how critical finite dimensional fluctuations affect the the Gardner transition found within mean-field theory
is no more an abstract and academic question. It has become a open and very relevant issue.
The aim of this work is addressing it using \emph{perturbative} renormalisation group techniques.  \\
Before starting our analysis there are two important points worth clarifying. The first question that comes to mind
when one applies mean-field results to estimate critical exponents in three dimensional systems concerns the role
of finite dimensional fluctuations. In this respect, it is important to stress 
that the exponents of the jamming transition found within mean-field theory are related to the soft-modes of the 
FRSB phase \cite{LNSW10, Wy12, WNW05, WSNW05, LDW13, BW06, BW07, BW09b}. In consequence, they are related to the properties of the low temperature/high pressure phase. They are not exponents related to a phase transition. There are therefore two separate issues: one is how the Gardner transition and its critical properties change from infinite dimension down to three, and the other is how the soft modes of the symmetry broken phase change from infinite dimension down to three. An instructive example is provided by the ferromagnetic Heisenberg model: its critical properties at the ferromagnetic phase transition change below four dimensions with respect to the mean-field ones, however the properties of the soft-Goldstone modes remain the same, as shown by analyzing the corresponding non-linear $\s$-model \cite{ZinnBook}. 
What we do in this work is to address the first issue, i.e. 
we focus on the critical properties of the Gardner transition. We shall just touch upon the second one in the conclusion.    \\
The other point we want to address is the relationship between the FRSB physics found for hard spheres in infinite dimensions and the one of spin-glasses in a field. In both cases one finds a FRSB phase without any residual 
symmetry present. Thus, reasoning only in terms of phases and type of symmetry breaking one would  
conclude that spin-glasses in a field and low temperature/high pressure glasses are in the same universality class both for the transition, Gardner versus spin-glass, and the properties of the FRSB phase. 
This would be also what one would conclude from the works by Moore and collaborators \cite{MY06, FM13, DM02}, in which  
the glass transition was argued to be related to the spin-glass transition in a field (see also \cite{Janus13}). 
As we shall show, however, the situation is more intricate and need further analysis. \\
The starting point of our derivation is the effective replica field theory which describes the critical fluctuations at the Gardner transition. At $T_G$ there is a phase transition from a 1RSB to a FRSB phase. Hence, the action of the theory can be formulated in terms of a fluctuating space dependent overlap field $Q_{ab}( r)=\phi_{ab}( r)+Q$ where $Q$ is the 1RSB value of the solution of the saddle point equations. The replica indices $a,b$ run from $1$ to $m$ and an analytic continuation for $m$ to real values is always assumed.
In our calculation $m$ is the 1RSB breaking point and must be considered fixed to the value reached at $T_G$ \cite{MPV87, footnotem}. 
There are no extra $n/m\rightarrow 0$ replicas since we focus on systems without quenched disorder \cite{footnoten}.
Other values of $m$, different from $m(T_G)$, can be used to select non-equilibrium metastable states within mean-field theory \cite{Mo95}. Whether our analysis for generic values of $m$ can be applied to metastable states will be discussed in the conclusion.   
In order to construct the most general action for $\phi_{ab}( r)$ we recall that there is no other symmetry that has to be taken into account besides replica permutation. As a consequence, one has to consider all quadratic and cubic terms 
allowed by replica symmetry. By analyzing the quadratic terms of the expansion of the action one recognizes that the replica field theory has a mass matrix that can be easily diagonalized \cite{TDP02}. Three distinct eigenvalues are found: the replicon, the longitudinal and the anomalous one. 
The mean-field analysis shows that the Gardner transition corresponds to the vanishing of the replicon eigenvalue whereas the others remain massive \cite{Ga85}. This means that only the replicon modes are critical and 
the others can be safely integrated out. Thus, in order to obtain the action of the critical modes we
only take into account the contribution from the critical replicon modes to the fluctuating overlap field. This is a standard procedure and it has been already followed in the case of the Edwards-Anderson (EA) model in a field \cite{BR80}. As expected, the results of the fixed points of the renormalization group equations are the same if the non critical modes are also taken into account \cite{TDP02b}.
The action that one obtains reads: \cite{TDP02b, BR80}
\beq
\begin{split}
\mathcal L&=\frac{1}{2}\sum_{\mathbf p}\left[(\mathbf p^2+r)\sum_{a,b=1}^{m}\phi_{ab}(\mathbf p)\phi_{ab}(-\mathbf p)\right]\\
&-\tilde{\sum}_{\mathbf p_1,\mathbf p_2,\mathbf p_3}\left[\frac 16g_1 \sum_{a,b,c}\phi_{ab}(\mathbf p_1)\phi_{bc}(\mathbf p_2)\phi_{ca}(\mathbf p_3)\right.\\
&\left.+\frac{1}{12}g_2\sum_{a,b}\phi_{ab}(\mathbf p_1)\phi_{ab}(\mathbf p_2)\phi_{ab}(\mathbf p_3)\right]
\end{split}
\eeq
where $\tilde \sum$ denotes a sum over momenta that sum up to zero in order to ensure translational invariance. 
The field $\phi_{ab}(\mathbf p)$ has the following properties
\beq
\begin{split}
&\phi_{ab}(\mathbf p)=\phi_{ba}(\mathbf p)\,\, , \,\, \phi_{aa}(\mathbf p)=0\\
&\sum_{b(\neq a)}\phi_{ab}(\mathbf p)=\sum_{a(\neq b)}\phi_{ab}(\mathbf p)=0
\end{split}
\eeq
that characterize the replicon eigenspace.
In the limit $m\to 0$ this replica field theory describes the behavior of the EA model in a field. To study the Gardner
transition we instead have to keep $m$ finite.  The perturbative renormalization group equations for such a theory 
were obtained in Ref. \cite{BR80, TDP02b} by performing the $\e$-expansion around $d=6$, which corresponds to
 the upper critical dimension of the theory (at least perturbatively, more on this later on). The equations that describe the RG flow
 read \cite{TDP02b}:
\beq\nonumber
\begin{split}
&\frac{\de r}{\de l}=(2-\frac 13\eta)r-\left[g_1^2\frac{m^4-8m^3+19m^2-4m-16}{(m-1)(m-2)^2}\right.\\
&\left.+g_1g_2\frac{2(3m^2-15m+16)}{(m-1)(m-2)^2}+g_2^2\frac{m^3-9m^2+26m-22}{2(m-1)(m-2)^2}\right]I_2\\
&\frac{\de g_1}{\de l}=\frac 12(\e -\eta)g_1+\left[A_1 g_1^3+A_2 g_1^2g_2+A_3g_1g_2^2+A_4g_2^3\right]I_3\\
&\frac{\de g_2}{\de l}=\frac 12(\e-\eta)g_2-\left[B_1 g_1^3+B_2 g_1^2g_2+B_3g_1g_2^2+B_4g_2^3\right]I_3
\end{split}
\eeq
where $I_2=(1+r)^{-2}$, $I_3=(1+r)^{-3}$, $\e=6-d$, $d$ is the spatial dimension, and $\eta$ the usual critical exponent related to the anomalous dimension of the field. The coefficients $A$ and $B$ reads:
\beq
\begin{split}
&A_1=\frac{m^5-10m^4+33m^3-8m^2-104m+112}{(m-1)(m-2)^3}\\
&A_2=\frac{3(3m^3-27m^2+64m-48)}{(m-1)(m-2)^3}\\
&A_3=\frac{3(-m^3+8m^2-17m+12)}{(m-1)(m-2)^3}\\
&A_4=-\frac{1}{(m-2)^3}\\
&B_1=\frac{24m}{(m-2)^2}\\
&B_2=\frac{6(m^3-5m^2-8m+16)}{(m-1)(m-2)^2}\\
&B_3=-\frac{3(6m^2-38m+40)}{(m-1)(m-2)^2}\\
&B_4=-\frac{m^3-11m^2+38m-34}{(m-1)(m-2)^2}
\end{split}
\eeq
Moreover we have
\beq
\begin{split}
\eta&=\left(H_1g_1^2+H_2g_1g_2+H_3g_2^2\right)\frac{1+r}{(1+r)^4}\\
H_1&=\frac{m^4-8m^3+19m^2-4m-16}{(m-1)(m-2)^2}\\
H_2&=\frac{2(3m^2-15m+16)}{(m-1)(m_2)^2}\\
H_3&=\frac{m^3-9m^2+26m-22}{2(m-1)(m-2)^2}\:.
\end{split}
\eeq
Starting from these equations we can write down an equation for $\l=g_2/g_1$ at criticality, i.e for $r=0$:
\beq \label{eql}
\begin{split}
\frac{\de \l}{\de l}&=-g_1^2\left[K_1+K_2\l+K_3\l^2+K_4\l^2+K_5\l^4\right]\\
K_1&=B_1\ \ \ \  K_5=A_4\ \ \ \ K_2=B_2+A_1\\
K_4&=B_4+A_3\ \ \ \ K_3=B_3+A_2\:.
\end{split}
\eeq
Equation (\ref{eql}) is dimension independent and thus it is particularly useful to discuss the RG fixed points (FP) \cite{footnotel}.
The fixed points equation for $\l$ reads
\beq \label{lambdaeq}
K_1+K_2\l+K_3\l^2+K_4\l^2+K_5\l^4=0
\eeq
where we consider $g_1 \neq 0$ since, as we verified, there are no RG-FPs characterized by $g_1=0$
except the Gaussian one for which $g_1=g_2=0$.
We find that that for $m<m^*\simeq 0.894$, eq. (\ref{lambdaeq}) has two real solutions $\l_1(m)$ and $\l_2(m)$.	
For $m\geq m^*$ other two real solutions  $\l_3(m)$ and $\l_4(m)$ appear.  
Note that for each value of $\l$ there are two RG-FPs related by the transformation $(g_1,g_2)\rightarrow (-g_1,-g_2)$. 
In the following we consider separately the $d>6$ and $d<6$ cases.
\begin{figure}
\scalebox{0.7}{
\input{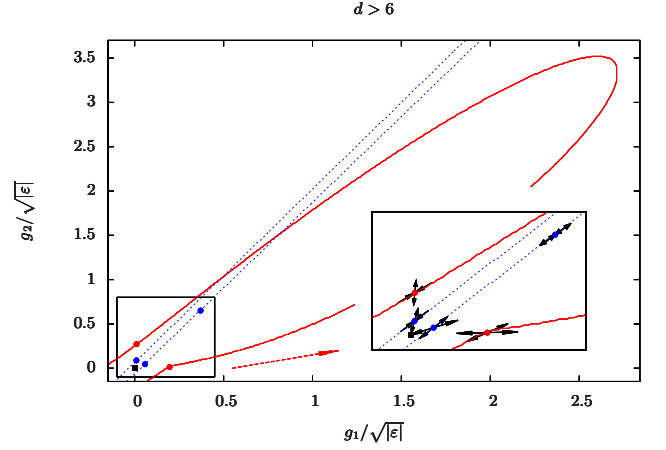}}
\caption{Basins of attraction of the Gaussian fixed point for $d>6$ in the 
quadrant $g_1>0,g_2>0$. The (blue) dots denote the non-Gaussian fixed points. 
The (red) continuous and (blue) dotted lines correspond respectively to the basin of attraction for $m=0.1\in [0,\tilde m]$ and $m=0.95\in [\tilde m, 1]$. We denote with arrows the stability of the FPs and the corresponding eigendirections (for some FPs these are almost collinear and not well distinguishable). The square denotes the stable Gaussian FP.   
}
\label{fig:dmaggiore6}
\end{figure}
For the former case we find that $\l_4(m)$ leads always to purely imaginary FPs and, hence, can be disregarded, whereas $\l_3(m)$ gives a real value for $g_1$ and $g_2$ for $m\geq \tilde m\simeq 0.905$ ($>m^*)$ only. Thus, depending whether 
$m\in[0,\tilde m]$ or $m\in[\tilde m,1]$ one finds four or six FPs (none of them stable). They all belong to the the border of the basin of attraction of the Gaussian fixed point (G-FP), see  Fig. \ref{fig:dmaggiore6}. Increasing the value of $m$ the basin of attraction stretches along the diagonal direction and becomes very large, or possibly infinite, for $m$ close to $\tilde m$. Its size shrinks to zero when $d\downarrow 6$, as found in Ref. \cite{BM11} in the case $m=0$. This is the 
prelude of what happens crossing $d=6$, where the G-FP becomes unstable.  
For $d<6$ the only physical (i.e. real) FPs are given by $\l_3(m)$ and $\l_4(m)$ for $m\in[m^*,\tilde m]$
and by $\l_4(m)$ for $m\in[m^*,1]$. We never find an attractive FPs except in the tiny
regime $m\in[m^*,\tilde m]$. See Fig. \ref{fig:dminore6} for a summary of the $d<6$ case.
We now discuss the main consequences of the perturbative RG results found above. 
The situation is very different from the one corresponding to standard field theories, e.g. the $\phi^4$ field theory,
because the basin of attraction of the G-FP shrinks to zero approaching $d=6$ and we do not find any attractive FP below (we neglect for the moment the case $m\in[m^*,\tilde m]$). Since the bare values of the coupling constants $g_1$ and $g_2$ are not arbitrarily small for a realistic system, the corresponding RG flow is bound to 
escape from the G-FP {\it strictly} before $d=6$ \cite{footnoted}.
This has two consequences: first, the upper critical dimension of the theory, $d_u$, above which mean-field results hold is strictly larger than six and, second, it is not universal since it depends on where the initial condition lies with respect to the basin of attraction of the G-FP. For $d<d_u$, the system flows to strong coupling, i.e. to a regime that we cannot access perturbatively. Different physical situations can correspond to this behaviour. The transition can be destroyed or can become first order for $d<d_u$ \cite{Cardybook,CCDDM05}. Another appealing possibility is that it remains critical but the critical behavior is dominated by a non-perturbative fixed point. For the spin-glass transition in a field or without time-reversal symmetry, which corresponds to the case $m=0$, the latter scenario is supported both by numerical simulations \cite{FP99,CCP98, PPR99,LKMY10, JanusPNAS, LPRR09} and real space RG analyses \cite{AB14} at least in high enough dimensions (the behaviour in three dimension is still controversial).
Note that although our results are overall very similar to the ones of spin-glasses in a field \cite{BR80, BM11}, the detailed behaviour of the RG flow in the perturbative regime is different \cite{footnoteG}. This, together with the fact for spin-glasses
that the critical fluctuating field has a number of components $m(m-3)/2$ with $m\to 0$ while in the case of the Gardner transition the number of replicas $m$ is fixed and positive, suggests that the universality classes should be distinct.  
On the other hand, if the theories at different $m$ have different critical behaviours then the critical properties of the Gardner transition are system-dependent, i.e. not universal at all (a quite weird physical situation), since the value of the breaking point $m$ at the transition is system-dependent. 
Non-perturbative RG treatments and numerical simulations are needed to clarify these issues.
The regime $m\in[m^*,\tilde m]$ that we neglected before is very peculiar since one does find an attractive FP below six dimensions but with a basin of attraction that shrinks to zero when $d\uparrow 6$. This leads to a rather baroque RG phenomenology. Since the interval $[m^*,\tilde m]$ 
is not only very tiny but also very close to one, it corresponds to systems (if any) extremely fine tuned. For this reason, 
we shall not address it further in this work.   \\
One of the main motivation to study the fate of the Gardner transition in finite dimensions is the recent discovery of the FRSB phase of hard spheres in high dimensions and its relevance for the properties of amorphous solids. 
As already discussed, within mean-field computations or in the limit of infinite dimensions, one can tune
the value of $m$ to select certain metastable states, e.g. corresponding to packings with a given volume fraction. 
The problem in applying our results to this case is that the procedure of selecting metastable states tuning $m$ is not well defined beyond mean-field 
theory. The 1RSB solution that is used in the mean-field computations is known to be unstable because of non-perturative
effects \cite{Fr05}, which simply correspond to the fact that the corresponding states are {\it meta}stable in any finite dimension. One way out of this problem is constraining the particles to only move around 
the positions they have in a given packing, as in the model introduced in Ref. \cite{FM13}. 
It is interesting to notice that this procedure explicitly introduces quenched disorder. Although we did not attempt to study this case by RG, 
we conjecture that a relationship with spin glasses in a field could emerge since
the disorder select a given metabasin transforming the Gardner transition we analyzed into a transition from a RS phase (describing the metabasin at high temperature/low pressure) to a FRSB phase (describing the multi-valley structure
inside the metabasin), similarly to what happens for the EA model in a field. Results supporting this view were 
presented in Ref. \cite{FM13, DM02}. This is certainly an issue worth investigating more both analytically and numerically. Establishing a direct relationship between the behavior of spin-glasses in a field and low temperature/high pressure glasses would be extremely important and useful, as argued in \cite{FM13, DM02}. Our \emph{perturbative} RG results cannot lead to any conclusive result on this. What they makes clear, however, is that the Gardner transition in three dimension either does not exists or it has a different nature from the mean-field one, namely it can become non-perturbative in the RG sense or first order. Numerical simulations and 
non-perturbative RG treatments are crucially needed in order to find out which one among these three possibilities is 
realized. 

\paragraph*{Acknowledgments --}{
We thank C. Cammarota, B. Delamotte, S. Franz, M. Moore, G. Parisi, G. Tarjus, M. Tarzia, F. Zamponi for useful discussions. We acknowledge financial support from the ERC grant NPRGGLASS.
}

\begin{thebibliography}{57}%
\makeatletter
\providecommand \@ifxundefined [1]{%
 \@ifx{#1\undefined}
}%
\providecommand \@ifnum [1]{%
 \ifnum #1\expandafter \@firstoftwo
 \else \expandafter \@secondoftwo
 \fi
}%
\providecommand \@ifx [1]{%
 \ifx #1\expandafter \@firstoftwo
 \else \expandafter \@secondoftwo
 \fi
}%
\providecommand \natexlab [1]{#1}%
\providecommand \enquote  [1]{``#1''}%
\providecommand \bibnamefont  [1]{#1}%
\providecommand \bibfnamefont [1]{#1}%
\providecommand \citenamefont [1]{#1}%
\providecommand \href@noop [0]{\@secondoftwo}%
\providecommand \href [0]{\begingroup \@sanitize@url \@href}%
\providecommand \@href[1]{\@@startlink{#1}\@@href}%
\providecommand \@@href[1]{\endgroup#1\@@endlink}%
\providecommand \@sanitize@url [0]{\catcode `\\12\catcode `\$12\catcode
  `\&12\catcode `\#12\catcode `\^12\catcode `\_12\catcode `\%12\relax}%
\providecommand \@@startlink[1]{}%
\providecommand \@@endlink[0]{}%
\providecommand \url  [0]{\begingroup\@sanitize@url \@url }%
\providecommand \@url [1]{\endgroup\@href {#1}{\urlprefix }}%
\providecommand \urlprefix  [0]{URL }%
\providecommand \Eprint [0]{\href }%
\providecommand \doibase [0]{http://dx.doi.org/}%
\providecommand \selectlanguage [0]{\@gobble}%
\providecommand \bibinfo  [0]{\@secondoftwo}%
\providecommand \bibfield  [0]{\@secondoftwo}%
\providecommand \translation [1]{[#1]}%
\providecommand \BibitemOpen [0]{}%
\providecommand \bibitemStop [0]{}%
\providecommand \bibitemNoStop [0]{.\EOS\space}%
\providecommand \EOS [0]{\spacefactor3000\relax}%
\providecommand \BibitemShut  [1]{\csname bibitem#1\endcsname}%
\let\auto@bib@innerbib\@empty
\bibitem [{\citenamefont {Kirkpatrick}\ and\ \citenamefont
  {Wolynes}(1987)}]{KW87}%
  \BibitemOpen
  \bibfield  {author} {\bibinfo {author} {\bibfnamefont {T.~R.}\ \bibnamefont
  {Kirkpatrick}}\ and\ \bibinfo {author} {\bibfnamefont {P.~G.}\ \bibnamefont
  {Wolynes}},\ }\href {\doibase 10.1103/PhysRevA.35.3072} {\bibfield  {journal}
  {\bibinfo  {journal} {Phys. Rev. A}\ }\textbf {\bibinfo {volume} {35}},\
  \bibinfo {pages} {3072} (\bibinfo {year} {1987})}\BibitemShut {NoStop}%
\bibitem [{\citenamefont {Kirkpatrick}\ and\ \citenamefont
  {Thirumalai}(1987)}]{KT87}%
  \BibitemOpen
  \bibfield  {author} {\bibinfo {author} {\bibfnamefont {T.~R.}\ \bibnamefont
  {Kirkpatrick}}\ and\ \bibinfo {author} {\bibfnamefont {D.}~\bibnamefont
  {Thirumalai}},\ }\href@noop {} {\bibfield  {journal} {\bibinfo  {journal}
  {Phys. Rev. Lett.}\ }\textbf {\bibinfo {volume} {58}},\ \bibinfo {pages}
  {2091} (\bibinfo {year} {1987})}\BibitemShut {NoStop}%
\bibitem [{\citenamefont {Kirkpatrick}\ \emph {et~al.}(1989)\citenamefont
  {Kirkpatrick}, \citenamefont {Thirumalai},\ and\ \citenamefont
  {Wolynes}}]{KTW89}%
  \BibitemOpen
  \bibfield  {author} {\bibinfo {author} {\bibfnamefont {T.~R.}\ \bibnamefont
  {Kirkpatrick}}, \bibinfo {author} {\bibfnamefont {D.}~\bibnamefont
  {Thirumalai}}, \ and\ \bibinfo {author} {\bibfnamefont {P.~G.}\ \bibnamefont
  {Wolynes}},\ }\href {\doibase 10.1103/PhysRevA.40.1045} {\bibfield  {journal}
  {\bibinfo  {journal} {Phys. Rev. A}\ }\textbf {\bibinfo {volume} {40}},\
  \bibinfo {pages} {1045} (\bibinfo {year} {1989})}\BibitemShut {NoStop}%
\bibitem [{\citenamefont {Biroli}\ and\ \citenamefont {Bouchaud}(2012)}]{BB09}%
  \BibitemOpen
  \bibfield  {author} {\bibinfo {author} {\bibfnamefont {G.}~\bibnamefont
  {Biroli}}\ and\ \bibinfo {author} {\bibfnamefont {J.}~\bibnamefont
  {Bouchaud}},\ }in\ \href@noop {} {\emph {\bibinfo {booktitle} {Structural
  Glasses and Supercooled Liquids: Theory, Experiment and Applications}}},\
  \bibinfo {editor} {edited by\ \bibinfo {editor} {\bibnamefont {P.G.Wolynes}}\
  and\ \bibinfo {editor} {\bibnamefont {V.Lubchenko}}}\ (\bibinfo  {publisher}
  {Wiley \& Sons},\ \bibinfo {year} {2012})\ \Eprint {http://arxiv.org/abs/{\tt
  arXiv:0912.2542}} {{\tt arXiv:0912.2542}} \BibitemShut {NoStop}%
\bibitem [{\citenamefont {Berthier}\ and\ \citenamefont {Biroli}(2011)}]{BB11}%
  \BibitemOpen
  \bibfield  {author} {\bibinfo {author} {\bibfnamefont {L.}~\bibnamefont
  {Berthier}}\ and\ \bibinfo {author} {\bibfnamefont {G.}~\bibnamefont
  {Biroli}},\ }\href {\doibase 10.1103/RevModPhys.83.587} {\bibfield  {journal}
  {\bibinfo  {journal} {Rev. Mod. Phys.}\ }\textbf {\bibinfo {volume} {83}},\
  \bibinfo {pages} {587} (\bibinfo {year} {2011})}\BibitemShut {NoStop}%
\bibitem [{\citenamefont {Wolynes}\ and\ \citenamefont
  {Lubchenko}(2012)}]{WL12}%
  \BibitemOpen
  \bibinfo {editor} {\bibfnamefont {P.}~\bibnamefont {Wolynes}}\ and\ \bibinfo
  {editor} {\bibfnamefont {V.}~\bibnamefont {Lubchenko}},\ eds.,\ \href@noop {}
  {\emph {\bibinfo {title} {Structural Glasses and Supercooled Liquids: Theory,
  Experiment, and Applications}}}\ (\bibinfo  {publisher} {Wiley},\ \bibinfo
  {year} {2012})\BibitemShut {NoStop}%
\bibitem [{\citenamefont {Kirkpatrick}\ and\ \citenamefont
  {Thirumalai}(2014)}]{KT14}%
  \BibitemOpen
  \bibfield  {author} {\bibinfo {author} {\bibfnamefont {T.}~\bibnamefont
  {Kirkpatrick}}\ and\ \bibinfo {author} {\bibfnamefont {D.}~\bibnamefont
  {Thirumalai}},\ }\href@noop {} {\bibfield  {journal} {\bibinfo  {journal}
  {arXiv preprint arXiv:1401.2024}\ } (\bibinfo {year} {2014})}\BibitemShut
  {NoStop}%
\bibitem [{\citenamefont {Krzakala}\ \emph {et~al.}(pear)\citenamefont
  {Krzakala}, \citenamefont {Ricci-Tersenghi}, \citenamefont {Zdeborova},
  \citenamefont {Zecchina}, \citenamefont {Tramel},\ and\ \citenamefont
  {Cugliandolo}}]{KrzakalaBook}%
  \BibitemOpen
  \bibinfo {editor} {\bibfnamefont {F.}~\bibnamefont {Krzakala}}, \bibinfo
  {editor} {\bibfnamefont {F.}~\bibnamefont {Ricci-Tersenghi}}, \bibinfo
  {editor} {\bibfnamefont {L.}~\bibnamefont {Zdeborova}}, \bibinfo {editor}
  {\bibfnamefont {R.}~\bibnamefont {Zecchina}}, \bibinfo {editor}
  {\bibfnamefont {E.~W.}\ \bibnamefont {Tramel}}, \ and\ \bibinfo {editor}
  {\bibfnamefont {L.~F.}\ \bibnamefont {Cugliandolo}},\ eds.,\ \href@noop {}
  {\emph {\bibinfo {title} {Statistical Physics, Optimization, Inference, and
  Message-Passing Algorithms}}}\ (\bibinfo {year} {To appear})\BibitemShut
  {NoStop}%
\bibitem [{\citenamefont {Kob}\ and\ \citenamefont {Berthier}(2013)}]{BK13}%
  \BibitemOpen
  \bibfield  {author} {\bibinfo {author} {\bibfnamefont {W.}~\bibnamefont
  {Kob}}\ and\ \bibinfo {author} {\bibfnamefont {L.}~\bibnamefont {Berthier}},\
  }\href@noop {} {\bibfield  {journal} {\bibinfo  {journal} {Physical Review
  Letters}\ }\textbf {\bibinfo {volume} {110}},\ \bibinfo {pages} {245702}
  (\bibinfo {year} {2013})}\BibitemShut {NoStop}%
\bibitem [{\citenamefont {Lubchenko}\ and\ \citenamefont
  {Wolynes}(2007)}]{LW07}%
  \BibitemOpen
  \bibfield  {author} {\bibinfo {author} {\bibfnamefont {V.}~\bibnamefont
  {Lubchenko}}\ and\ \bibinfo {author} {\bibfnamefont {P.~G.}\ \bibnamefont
  {Wolynes}},\ }\href {\doibase 10.1146/annurev.physchem.58.032806.104653}
  {\bibfield  {journal} {\bibinfo  {journal} {Annual Review of Physical
  Chemistry}\ }\textbf {\bibinfo {volume} {58}},\ \bibinfo {pages} {235}
  (\bibinfo {year} {2007})}\BibitemShut {NoStop}%
\bibitem [{\citenamefont {Biroli}(2014)}]{Bi14}%
  \BibitemOpen
  \bibfield  {author} {\bibinfo {author} {\bibfnamefont {G.}~\bibnamefont
  {Biroli}},\ }\href@noop {} {\bibfield  {journal} {\bibinfo  {journal} {Nature
  Physics}\ }\textbf {\bibinfo {volume} {10}},\ \bibinfo {pages} {555}
  (\bibinfo {year} {2014})}\BibitemShut {NoStop}%
\bibitem [{\citenamefont {Liu}\ and\ \citenamefont {Nagel}(1998)}]{LN98}%
  \BibitemOpen
  \bibfield  {author} {\bibinfo {author} {\bibfnamefont {A.~J.}\ \bibnamefont
  {Liu}}\ and\ \bibinfo {author} {\bibfnamefont {S.~R.}\ \bibnamefont
  {Nagel}},\ }\href@noop {} {\bibfield  {journal} {\bibinfo  {journal}
  {Nature}\ }\textbf {\bibinfo {volume} {396}},\ \bibinfo {pages} {21}
  (\bibinfo {year} {1998})}\BibitemShut {NoStop}%
\bibitem [{\citenamefont {Liu}\ \emph {et~al.}(2011)\citenamefont {Liu},
  \citenamefont {Nagel}, \citenamefont {Van~Saarloos},\ and\ \citenamefont
  {Wyart}}]{LNSW10}%
  \BibitemOpen
  \bibfield  {author} {\bibinfo {author} {\bibfnamefont {A.}~\bibnamefont
  {Liu}}, \bibinfo {author} {\bibfnamefont {S.}~\bibnamefont {Nagel}}, \bibinfo
  {author} {\bibfnamefont {W.}~\bibnamefont {Van~Saarloos}}, \ and\ \bibinfo
  {author} {\bibfnamefont {M.}~\bibnamefont {Wyart}},\ }in\ \href@noop {}
  {\emph {\bibinfo {booktitle} {Dynamical Heterogeneities and Glasses}}},\
  \bibinfo {editor} {edited by\ \bibinfo {editor} {\bibfnamefont
  {L.}~\bibnamefont {Berthier}}, \bibinfo {editor} {\bibfnamefont
  {G.}~\bibnamefont {Biroli}}, \bibinfo {editor} {\bibfnamefont {J.-P.}\
  \bibnamefont {Bouchaud}}, \bibinfo {editor} {\bibfnamefont {L.}~\bibnamefont
  {Cipelletti}}, \ and\ \bibinfo {editor} {\bibfnamefont {W.}~\bibnamefont {van
  Saarloos}}}\ (\bibinfo  {publisher} {Oxford University Press},\ \bibinfo
  {year} {2011})\ \Eprint {http://arxiv.org/abs/{\tt arXiv:1006.2365}} {{\tt
  arXiv:1006.2365}} \BibitemShut {NoStop}%
\bibitem [{\citenamefont {Charbonneau}\ \emph
  {et~al.}(2014{\natexlab{a}})\citenamefont {Charbonneau}, \citenamefont
  {Kurchan}, \citenamefont {Parisi}, \citenamefont {Urbani},\ and\
  \citenamefont {Zamponi}}]{CKPUZ14}%
  \BibitemOpen
  \bibfield  {author} {\bibinfo {author} {\bibfnamefont {P.}~\bibnamefont
  {Charbonneau}}, \bibinfo {author} {\bibfnamefont {J.}~\bibnamefont
  {Kurchan}}, \bibinfo {author} {\bibfnamefont {G.}~\bibnamefont {Parisi}},
  \bibinfo {author} {\bibfnamefont {P.}~\bibnamefont {Urbani}}, \ and\ \bibinfo
  {author} {\bibfnamefont {F.}~\bibnamefont {Zamponi}},\ }\href@noop {}
  {\bibfield  {journal} {\bibinfo  {journal} {Nature Communications}\ }\textbf
  {\bibinfo {volume} {5}},\ \bibinfo {pages} {3725} (\bibinfo {year}
  {2014}{\natexlab{a}})}\BibitemShut {NoStop}%
\bibitem [{\citenamefont {Gross}\ \emph {et~al.}(1985)\citenamefont {Gross},
  \citenamefont {Kanter},\ and\ \citenamefont {Sompolinsky}}]{GKS85}%
  \BibitemOpen
  \bibfield  {author} {\bibinfo {author} {\bibfnamefont {D.~J.}\ \bibnamefont
  {Gross}}, \bibinfo {author} {\bibfnamefont {I.}~\bibnamefont {Kanter}}, \
  and\ \bibinfo {author} {\bibfnamefont {H.}~\bibnamefont {Sompolinsky}},\
  }\href@noop {} {\bibfield  {journal} {\bibinfo  {journal} {Physical Review
  Letters}\ }\textbf {\bibinfo {volume} {55}} (\bibinfo {year}
  {1985})}\BibitemShut {NoStop}%
\bibitem [{\citenamefont {Gardner}(1985)}]{Ga85}%
  \BibitemOpen
  \bibfield  {author} {\bibinfo {author} {\bibfnamefont {E.}~\bibnamefont
  {Gardner}},\ }\href@noop {} {\bibfield  {journal} {\bibinfo  {journal}
  {Nuclear Physics B}\ }\textbf {\bibinfo {volume} {257}},\ \bibinfo {pages}
  {747} (\bibinfo {year} {1985})}\BibitemShut {NoStop}%
\bibitem [{\citenamefont {Gross}\ and\ \citenamefont {M\'ezard}(1984)}]{GM84}%
  \BibitemOpen
  \bibfield  {author} {\bibinfo {author} {\bibfnamefont {D.~J.}\ \bibnamefont
  {Gross}}\ and\ \bibinfo {author} {\bibfnamefont {M.}~\bibnamefont
  {M\'ezard}},\ }\href@noop {} {\bibfield  {journal} {\bibinfo  {journal}
  {Nucl.~Phys.~B}\ }\textbf {\bibinfo {volume} {240}},\ \bibinfo {pages} {431}
  (\bibinfo {year} {1984})}\BibitemShut {NoStop}%
\bibitem [{\citenamefont {M\'ezard}\ \emph {et~al.}(1987)\citenamefont
  {M\'ezard}, \citenamefont {Parisi},\ and\ \citenamefont {Virasoro}}]{MPV87}%
  \BibitemOpen
  \bibfield  {author} {\bibinfo {author} {\bibfnamefont {M.}~\bibnamefont
  {M\'ezard}}, \bibinfo {author} {\bibfnamefont {G.}~\bibnamefont {Parisi}}, \
  and\ \bibinfo {author} {\bibfnamefont {M.~A.}\ \bibnamefont {Virasoro}},\
  }\href@noop {} {\emph {\bibinfo {title} {Spin glass theory and beyond}}}\
  (\bibinfo  {publisher} {World Scientific},\ \bibinfo {address} {Singapore},\
  \bibinfo {year} {1987})\BibitemShut {NoStop}%
\bibitem [{\citenamefont {Barrat}\ \emph {et~al.}(1997)\citenamefont {Barrat},
  \citenamefont {Franz},\ and\ \citenamefont {Parisi}}]{BFP97}%
  \BibitemOpen
  \bibfield  {author} {\bibinfo {author} {\bibfnamefont {A.}~\bibnamefont
  {Barrat}}, \bibinfo {author} {\bibfnamefont {S.}~\bibnamefont {Franz}}, \
  and\ \bibinfo {author} {\bibfnamefont {G.}~\bibnamefont {Parisi}},\ }\href
  {http://stacks.iop.org/0305-4470/30/5593} {\bibfield  {journal} {\bibinfo
  {journal} {Journal of Physics A: Mathematical and General}\ }\textbf
  {\bibinfo {volume} {30}},\ \bibinfo {pages} {5593} (\bibinfo {year}
  {1997})}\BibitemShut {NoStop}%
\bibitem [{\citenamefont {Montanari}\ and\ \citenamefont
  {Ricci-Tersenghi}(2003)}]{MR03}%
  \BibitemOpen
  \bibfield  {author} {\bibinfo {author} {\bibfnamefont {A.}~\bibnamefont
  {Montanari}}\ and\ \bibinfo {author} {\bibfnamefont {F.}~\bibnamefont
  {Ricci-Tersenghi}},\ }\href@noop {} {\bibfield  {journal} {\bibinfo
  {journal} {The European Physical Journal B-Condensed Matter and Complex
  Systems}\ }\textbf {\bibinfo {volume} {33}},\ \bibinfo {pages} {339}
  (\bibinfo {year} {2003})}\BibitemShut {NoStop}%
\bibitem [{\citenamefont {Rizzo}(2013)}]{Ri13}%
  \BibitemOpen
  \bibfield  {author} {\bibinfo {author} {\bibfnamefont {T.}~\bibnamefont
  {Rizzo}},\ }\href@noop {} {\bibfield  {journal} {\bibinfo  {journal}
  {Physical Review E}\ }\textbf {\bibinfo {volume} {88}},\ \bibinfo {pages}
  {032135} (\bibinfo {year} {2013})}\BibitemShut {NoStop}%
\bibitem [{\citenamefont {Kurchan}\ \emph {et~al.}(2012)\citenamefont
  {Kurchan}, \citenamefont {Parisi},\ and\ \citenamefont {Zamponi}}]{KPZ12}%
  \BibitemOpen
  \bibfield  {author} {\bibinfo {author} {\bibfnamefont {J.}~\bibnamefont
  {Kurchan}}, \bibinfo {author} {\bibfnamefont {G.}~\bibnamefont {Parisi}}, \
  and\ \bibinfo {author} {\bibfnamefont {F.}~\bibnamefont {Zamponi}},\
  }\href@noop {} {\bibfield  {journal} {\bibinfo  {journal} {Journal of
  Statistical Mechanics: Theory and Experiment}\ }\textbf {\bibinfo {volume}
  {2012}},\ \bibinfo {pages} {P10012} (\bibinfo {year} {2012})}\BibitemShut
  {NoStop}%
\bibitem [{\citenamefont {Kurchan}\ \emph {et~al.}(2013)\citenamefont
  {Kurchan}, \citenamefont {Parisi}, \citenamefont {Urbani},\ and\
  \citenamefont {Zamponi}}]{KPUZ13}%
  \BibitemOpen
  \bibfield  {author} {\bibinfo {author} {\bibfnamefont {J.}~\bibnamefont
  {Kurchan}}, \bibinfo {author} {\bibfnamefont {G.}~\bibnamefont {Parisi}},
  \bibinfo {author} {\bibfnamefont {P.}~\bibnamefont {Urbani}}, \ and\ \bibinfo
  {author} {\bibfnamefont {F.}~\bibnamefont {Zamponi}},\ }\href@noop {}
  {\bibfield  {journal} {\bibinfo  {journal} {The Journal of Physical Chemistry
  B}\ }\textbf {\bibinfo {volume} {117}},\ \bibinfo {pages} {12979} (\bibinfo
  {year} {2013})}\BibitemShut {NoStop}%
\bibitem [{\citenamefont {Charbonneau}\ \emph
  {et~al.}(2014{\natexlab{b}})\citenamefont {Charbonneau}, \citenamefont
  {Kurchan}, \citenamefont {Parisi}, \citenamefont {Urbani},\ and\
  \citenamefont {Zamponi}}]{jstatCKPUZ13}%
  \BibitemOpen
  \bibfield  {author} {\bibinfo {author} {\bibfnamefont {P.}~\bibnamefont
  {Charbonneau}}, \bibinfo {author} {\bibfnamefont {J.}~\bibnamefont
  {Kurchan}}, \bibinfo {author} {\bibfnamefont {G.}~\bibnamefont {Parisi}},
  \bibinfo {author} {\bibfnamefont {P.}~\bibnamefont {Urbani}}, \ and\ \bibinfo
  {author} {\bibfnamefont {F.}~\bibnamefont {Zamponi}},\ }\href@noop {}
  {\bibfield  {journal} {\bibinfo  {journal} {Journal of Statistical Mechanics:
  Theory and Experiment}\ }\textbf {\bibinfo {volume} {2014}},\ \bibinfo
  {pages} {P10009} (\bibinfo {year} {2014}{\natexlab{b}})}\BibitemShut
  {NoStop}%
\bibitem [{\citenamefont {Wyart}(2012)}]{Wy12}%
  \BibitemOpen
  \bibfield  {author} {\bibinfo {author} {\bibfnamefont {M.}~\bibnamefont
  {Wyart}},\ }\href@noop {} {\bibfield  {journal} {\bibinfo  {journal} {Phys.
  Rev. Lett.}\ }\textbf {\bibinfo {volume} {109}},\ \bibinfo {pages} {125502}
  (\bibinfo {year} {2012})}\BibitemShut {NoStop}%
\bibitem [{\citenamefont {Wyart}\ \emph
  {et~al.}(2005{\natexlab{a}})\citenamefont {Wyart}, \citenamefont {Nagel},\
  and\ \citenamefont {Witten}}]{WNW05}%
  \BibitemOpen
  \bibfield  {author} {\bibinfo {author} {\bibfnamefont {M.}~\bibnamefont
  {Wyart}}, \bibinfo {author} {\bibfnamefont {S.}~\bibnamefont {Nagel}}, \ and\
  \bibinfo {author} {\bibfnamefont {T.}~\bibnamefont {Witten}},\ }\href@noop {}
  {\bibfield  {journal} {\bibinfo  {journal} {Europhysics Letters}\ }\textbf
  {\bibinfo {volume} {72}},\ \bibinfo {pages} {486} (\bibinfo {year}
  {2005}{\natexlab{a}})}\BibitemShut {NoStop}%
\bibitem [{\citenamefont {Wyart}\ \emph
  {et~al.}(2005{\natexlab{b}})\citenamefont {Wyart}, \citenamefont {Silbert},
  \citenamefont {Nagel},\ and\ \citenamefont {Witten}}]{WSNW05}%
  \BibitemOpen
  \bibfield  {author} {\bibinfo {author} {\bibfnamefont {M.}~\bibnamefont
  {Wyart}}, \bibinfo {author} {\bibfnamefont {L.}~\bibnamefont {Silbert}},
  \bibinfo {author} {\bibfnamefont {S.}~\bibnamefont {Nagel}}, \ and\ \bibinfo
  {author} {\bibfnamefont {T.}~\bibnamefont {Witten}},\ }\href@noop {}
  {\bibfield  {journal} {\bibinfo  {journal} {Physical Review E}\ }\textbf
  {\bibinfo {volume} {72}},\ \bibinfo {pages} {051306} (\bibinfo {year}
  {2005}{\natexlab{b}})}\BibitemShut {NoStop}%
\bibitem [{\citenamefont {Lerner}\ \emph {et~al.}(2013)\citenamefont {Lerner},
  \citenamefont {During},\ and\ \citenamefont {Wyart}}]{LDW13}%
  \BibitemOpen
  \bibfield  {author} {\bibinfo {author} {\bibfnamefont {E.}~\bibnamefont
  {Lerner}}, \bibinfo {author} {\bibfnamefont {G.}~\bibnamefont {During}}, \
  and\ \bibinfo {author} {\bibfnamefont {M.}~\bibnamefont {Wyart}},\
  }\href@noop {} {\bibfield  {journal} {\bibinfo  {journal} {Soft Matter}\
  }\textbf {\bibinfo {volume} {9}},\ \bibinfo {pages} {8252} (\bibinfo {year}
  {2013})}\BibitemShut {NoStop}%
\bibitem [{\citenamefont {Brito}\ and\ \citenamefont {Wyart}(2006)}]{BW06}%
  \BibitemOpen
  \bibfield  {author} {\bibinfo {author} {\bibfnamefont {C.}~\bibnamefont
  {Brito}}\ and\ \bibinfo {author} {\bibfnamefont {M.}~\bibnamefont {Wyart}},\
  }\href {http://stacks.iop.org/0295-5075/76/149} {\bibfield  {journal}
  {\bibinfo  {journal} {Europhysics Letters (EPL)}\ }\textbf {\bibinfo {volume}
  {76}},\ \bibinfo {pages} {149} (\bibinfo {year} {2006})}\BibitemShut
  {NoStop}%
\bibitem [{\citenamefont {Brito}\ and\ \citenamefont {Wyart}(2007)}]{BW07}%
  \BibitemOpen
  \bibfield  {author} {\bibinfo {author} {\bibfnamefont {C.}~\bibnamefont
  {Brito}}\ and\ \bibinfo {author} {\bibfnamefont {M.}~\bibnamefont {Wyart}},\
  }\href@noop {} {\bibfield  {journal} {\bibinfo  {journal} {Journal of
  Statistical Mechanics: Theory and Experiment}\ }\textbf {\bibinfo {volume}
  {2007}},\ \bibinfo {pages} {L08003} (\bibinfo {year} {2007})}\BibitemShut
  {NoStop}%
\bibitem [{\citenamefont {Brito}\ and\ \citenamefont {Wyart}(2009)}]{BW09b}%
  \BibitemOpen
  \bibfield  {author} {\bibinfo {author} {\bibfnamefont {C.}~\bibnamefont
  {Brito}}\ and\ \bibinfo {author} {\bibfnamefont {M.}~\bibnamefont {Wyart}},\
  }\href@noop {} {\bibfield  {journal} {\bibinfo  {journal} {The Journal of
  Chemical Physics}\ }\textbf {\bibinfo {volume} {131}},\ \bibinfo {pages}
  {024504} (\bibinfo {year} {2009})}\BibitemShut {NoStop}%
\bibitem [{\citenamefont {Zinn-Justin}(1999)}]{ZinnBook}%
  \BibitemOpen
  \bibfield  {author} {\bibinfo {author} {\bibfnamefont {J.}~\bibnamefont
  {Zinn-Justin}},\ }\href@noop {} {\emph {\bibinfo {title} {Quantum Field
  Theory and Critical Phenomena}}}\ (\bibinfo  {publisher} {Oxford University
  Press},\ \bibinfo {year} {1999})\BibitemShut {NoStop}%
\bibitem [{\citenamefont {Moore}\ and\ \citenamefont {Yeo}(2006)}]{MY06}%
  \BibitemOpen
  \bibfield  {author} {\bibinfo {author} {\bibfnamefont {M.}~\bibnamefont
  {Moore}}\ and\ \bibinfo {author} {\bibfnamefont {J.}~\bibnamefont {Yeo}},\
  }\href@noop {} {\bibfield  {journal} {\bibinfo  {journal} {Physical Review
  Letters}\ }\textbf {\bibinfo {volume} {96}},\ \bibinfo {pages} {095701}
  (\bibinfo {year} {2006})}\BibitemShut {NoStop}%
\bibitem [{\citenamefont {Fullerton}\ and\ \citenamefont {Moore}(2013)}]{FM13}%
  \BibitemOpen
  \bibfield  {author} {\bibinfo {author} {\bibfnamefont {C.~J.}\ \bibnamefont
  {Fullerton}}\ and\ \bibinfo {author} {\bibfnamefont {M.}~\bibnamefont
  {Moore}},\ }\href@noop {} {\bibfield  {journal} {\bibinfo  {journal} {arXiv
  preprint arXiv:1304.4420}\ } (\bibinfo {year} {2013})}\BibitemShut {NoStop}%
\bibitem [{\citenamefont {Moore}\ and\ \citenamefont {Drossel}(2002)}]{DM02}%
  \BibitemOpen
  \bibfield  {author} {\bibinfo {author} {\bibfnamefont {M.}~\bibnamefont
  {Moore}}\ and\ \bibinfo {author} {\bibfnamefont {B.}~\bibnamefont
  {Drossel}},\ }\href@noop {} {\bibfield  {journal} {\bibinfo  {journal}
  {Physical Review Letters}\ }\textbf {\bibinfo {volume} {89}},\ \bibinfo
  {pages} {217202} (\bibinfo {year} {2002})}\BibitemShut {NoStop}%
\bibitem [{\citenamefont {Baity-Jesi}\ \emph {et~al.}(2013)\citenamefont
  {Baity-Jesi}, \citenamefont {Ba{\~n}os}, \citenamefont {Cruz}, \citenamefont
  {Fernandez}, \citenamefont {Gil-Narvion}, \citenamefont {I{\~n}iguez},
  \citenamefont {Maiorano}, \citenamefont {Mantovani}, \citenamefont
  {Marinari}, \citenamefont {Martin-Mayor} \emph {et~al.}}]{Janus13}%
  \BibitemOpen
  \bibfield  {author} {\bibinfo {author} {\bibfnamefont {M.}~\bibnamefont
  {Baity-Jesi}}, \bibinfo {author} {\bibfnamefont {R.~A.}\ \bibnamefont
  {Ba{\~n}os}}, \bibinfo {author} {\bibfnamefont {A.}~\bibnamefont {Cruz}},
  \bibinfo {author} {\bibfnamefont {L.}~\bibnamefont {Fernandez}}, \bibinfo
  {author} {\bibfnamefont {J.}~\bibnamefont {Gil-Narvion}}, \bibinfo {author}
  {\bibfnamefont {D.}~\bibnamefont {I{\~n}iguez}}, \bibinfo {author}
  {\bibfnamefont {A.}~\bibnamefont {Maiorano}}, \bibinfo {author}
  {\bibfnamefont {F.}~\bibnamefont {Mantovani}}, \bibinfo {author}
  {\bibfnamefont {E.}~\bibnamefont {Marinari}}, \bibinfo {author}
  {\bibfnamefont {V.}~\bibnamefont {Martin-Mayor}},  \emph {et~al.},\
  }\href@noop {} {\bibfield  {journal} {\bibinfo  {journal} {arXiv preprint
  arXiv:1307.4998}\ } (\bibinfo {year} {2013})}\BibitemShut {NoStop}%
\bibitem [{foo({\natexlab{a}})}]{footnotem}%
  \BibitemOpen
  \href@noop {} {\emph {\bibinfo {title} {\emph{The value of the breaking point
  parameter is chosen as the value that makes stationary the free energy, hence
  it is a microscopic non-universal quantity}}}}\BibitemShut {NoStop}%
\bibitem [{foo({\natexlab{b}})}]{footnoten}%
  \BibitemOpen
  \href@noop {} {\emph {\bibinfo {title} {\emph{When there is quenched disorder
  the total number of replicas is $n\to 0$. In this case our procedure of of
  focussing on $m$ replicas corresponds to taking into account fluctuations
  within metabasins and neglecting all the others}}}}\BibitemShut {NoStop}%
\bibitem [{\citenamefont {Monasson}(1995)}]{Mo95}%
  \BibitemOpen
  \bibfield  {author} {\bibinfo {author} {\bibfnamefont {R.}~\bibnamefont
  {Monasson}},\ }\href {\doibase 10.1103/PhysRevLett.75.2847} {\bibfield
  {journal} {\bibinfo  {journal} {Phys. Rev. Lett.}\ }\textbf {\bibinfo
  {volume} {75}},\ \bibinfo {pages} {2847} (\bibinfo {year}
  {1995})}\BibitemShut {NoStop}%
\bibitem [{\citenamefont {Temesv{\'a}ri}\ \emph {et~al.}(2002)\citenamefont
  {Temesv{\'a}ri}, \citenamefont {De~Dominicis},\ and\ \citenamefont
  {Pimentel}}]{TDP02}%
  \BibitemOpen
  \bibfield  {author} {\bibinfo {author} {\bibfnamefont {T.}~\bibnamefont
  {Temesv{\'a}ri}}, \bibinfo {author} {\bibfnamefont {C.}~\bibnamefont
  {De~Dominicis}}, \ and\ \bibinfo {author} {\bibfnamefont {I.}~\bibnamefont
  {Pimentel}},\ }\href@noop {} {\bibfield  {journal} {\bibinfo  {journal} {The
  European Physical Journal B-Condensed Matter and Complex Systems}\ }\textbf
  {\bibinfo {volume} {25}},\ \bibinfo {pages} {361} (\bibinfo {year}
  {2002})}\BibitemShut {NoStop}%
\bibitem [{\citenamefont {Bray}\ and\ \citenamefont {Roberts}(1980)}]{BR80}%
  \BibitemOpen
  \bibfield  {author} {\bibinfo {author} {\bibfnamefont {A.}~\bibnamefont
  {Bray}}\ and\ \bibinfo {author} {\bibfnamefont {S.}~\bibnamefont {Roberts}},\
  }\href@noop {} {\bibfield  {journal} {\bibinfo  {journal} {Journal of Physics
  C: Solid State Physics}\ }\textbf {\bibinfo {volume} {13}},\ \bibinfo {pages}
  {5405} (\bibinfo {year} {1980})}\BibitemShut {NoStop}%
\bibitem [{\citenamefont {Pimentel}\ \emph {et~al.}(2002)\citenamefont
  {Pimentel}, \citenamefont {Temesv{\'a}ri},\ and\ \citenamefont
  {De~Dominicis}}]{TDP02b}%
  \BibitemOpen
  \bibfield  {author} {\bibinfo {author} {\bibfnamefont {I.}~\bibnamefont
  {Pimentel}}, \bibinfo {author} {\bibfnamefont {T.}~\bibnamefont
  {Temesv{\'a}ri}}, \ and\ \bibinfo {author} {\bibfnamefont {C.}~\bibnamefont
  {De~Dominicis}},\ }\href@noop {} {\bibfield  {journal} {\bibinfo  {journal}
  {Physical Review B}\ }\textbf {\bibinfo {volume} {65}},\ \bibinfo {pages}
  {224420} (\bibinfo {year} {2002})}\BibitemShut {NoStop}%
\bibitem [{foo({\natexlab{c}})}]{footnotel}%
  \BibitemOpen
  \href@noop {} {\emph {\bibinfo {title} {\emph{The $\l$ of the exact Landau
  free energy, also called effective action, is connected to the perturbative
  2-RSB breaking point infinitesimally below the Gardner transition and, in
  order to have a physical fullRSB solution, it must verify $2m<\l<2$
  \cite{Ri13}. This $\l$ however differs from the one of the running action we
  are considering. In consequence one cannot use the previous inequality to
  restrict the possible value of RG fixed points}}}}\BibitemShut {NoStop}%
\bibitem [{\citenamefont {Moore}\ and\ \citenamefont {Bray}(2011)}]{BM11}%
  \BibitemOpen
  \bibfield  {author} {\bibinfo {author} {\bibfnamefont {M.}~\bibnamefont
  {Moore}}\ and\ \bibinfo {author} {\bibfnamefont {A.}~\bibnamefont {Bray}},\
  }\href@noop {} {\bibfield  {journal} {\bibinfo  {journal} {Physical Review
  B}\ }\textbf {\bibinfo {volume} {83}},\ \bibinfo {pages} {224408} (\bibinfo
  {year} {2011})}\BibitemShut {NoStop}%
\bibitem [{foo({\natexlab{d}})}]{footnoted}%
  \BibitemOpen
  \href@noop {} {\emph {\bibinfo {title} {\emph{The only way to avoid this
  conclusion and find $d_u=6$ is that there exists a bottleneck-effect for the
  non-perturbative RG flow, i.e. that: (1) there is a {\it finite}
  non-perturbative region in the space of coupling constants that flows under
  RG toward the arbitrary small (when $d\to 6^+$) basin of attraction in the
  $g_1, g_2$ plane; (2) all physical initial conditions belong to this region.
  We have no reason to believe that these two rather exceptional conditions
  hold (especially the second one)}}}}\BibitemShut {NoStop}%
\bibitem [{\citenamefont {Cardy}(1996)}]{Cardybook}%
  \BibitemOpen
  \bibfield  {author} {\bibinfo {author} {\bibfnamefont {J.}~\bibnamefont
  {Cardy}},\ }\href@noop {} {\emph {\bibinfo {title} {Scaling and
  renormalization in statistical physics}}},\ Vol.~\bibinfo {volume} {5}\
  (\bibinfo  {publisher} {Cambridge University Press},\ \bibinfo {year}
  {1996})\BibitemShut {NoStop}%
\bibitem [{\citenamefont {Canet}\ \emph {et~al.}(2005)\citenamefont {Canet},
  \citenamefont {Chat{\'e}}, \citenamefont {Delamotte}, \citenamefont
  {Dornic},\ and\ \citenamefont {Munoz}}]{CCDDM05}%
  \BibitemOpen
  \bibfield  {author} {\bibinfo {author} {\bibfnamefont {L.}~\bibnamefont
  {Canet}}, \bibinfo {author} {\bibfnamefont {H.}~\bibnamefont {Chat{\'e}}},
  \bibinfo {author} {\bibfnamefont {B.}~\bibnamefont {Delamotte}}, \bibinfo
  {author} {\bibfnamefont {I.}~\bibnamefont {Dornic}}, \ and\ \bibinfo {author}
  {\bibfnamefont {M.~A.}\ \bibnamefont {Munoz}},\ }\href@noop {} {\bibfield
  {journal} {\bibinfo  {journal} {Physical review letters}\ }\textbf {\bibinfo
  {volume} {95}},\ \bibinfo {pages} {100601} (\bibinfo {year}
  {2005})}\BibitemShut {NoStop}%
\bibitem [{\citenamefont {Franz}\ and\ \citenamefont {Parisi}(1999)}]{FP99}%
  \BibitemOpen
  \bibfield  {author} {\bibinfo {author} {\bibfnamefont {S.}~\bibnamefont
  {Franz}}\ and\ \bibinfo {author} {\bibfnamefont {G.}~\bibnamefont {Parisi}},\
  }\href@noop {} {\bibfield  {journal} {\bibinfo  {journal} {The European
  Physical Journal B-Condensed Matter and Complex Systems}\ }\textbf {\bibinfo
  {volume} {8}},\ \bibinfo {pages} {417} (\bibinfo {year} {1999})}\BibitemShut
  {NoStop}%
\bibitem [{\citenamefont {Campellone}\ \emph {et~al.}(1998)\citenamefont
  {Campellone}, \citenamefont {Coluzzi},\ and\ \citenamefont {Parisi}}]{CCP98}%
  \BibitemOpen
  \bibfield  {author} {\bibinfo {author} {\bibfnamefont {M.}~\bibnamefont
  {Campellone}}, \bibinfo {author} {\bibfnamefont {B.}~\bibnamefont {Coluzzi}},
  \ and\ \bibinfo {author} {\bibfnamefont {G.}~\bibnamefont {Parisi}},\
  }\href@noop {} {\bibfield  {journal} {\bibinfo  {journal} {Physical Review
  B}\ }\textbf {\bibinfo {volume} {58}},\ \bibinfo {pages} {12081} (\bibinfo
  {year} {1998})}\BibitemShut {NoStop}%
\bibitem [{\citenamefont {Parisi}\ \emph {et~al.}(1999)\citenamefont {Parisi},
  \citenamefont {Picco},\ and\ \citenamefont {Ritort}}]{PPR99}%
  \BibitemOpen
  \bibfield  {author} {\bibinfo {author} {\bibfnamefont {G.}~\bibnamefont
  {Parisi}}, \bibinfo {author} {\bibfnamefont {M.}~\bibnamefont {Picco}}, \
  and\ \bibinfo {author} {\bibfnamefont {F.}~\bibnamefont {Ritort}},\
  }\href@noop {} {\bibfield  {journal} {\bibinfo  {journal} {Physical Review
  E}\ }\textbf {\bibinfo {volume} {60}},\ \bibinfo {pages} {58} (\bibinfo
  {year} {1999})}\BibitemShut {NoStop}%
\bibitem [{\citenamefont {Larson}\ \emph {et~al.}(2010)\citenamefont {Larson},
  \citenamefont {Katzgraber}, \citenamefont {Moore},\ and\ \citenamefont
  {Young}}]{LKMY10}%
  \BibitemOpen
  \bibfield  {author} {\bibinfo {author} {\bibfnamefont {D.}~\bibnamefont
  {Larson}}, \bibinfo {author} {\bibfnamefont {H.~G.}\ \bibnamefont
  {Katzgraber}}, \bibinfo {author} {\bibfnamefont {M.}~\bibnamefont {Moore}}, \
  and\ \bibinfo {author} {\bibfnamefont {A.}~\bibnamefont {Young}},\
  }\href@noop {} {\bibfield  {journal} {\bibinfo  {journal} {Physical Review
  B}\ }\textbf {\bibinfo {volume} {81}},\ \bibinfo {pages} {064415} (\bibinfo
  {year} {2010})}\BibitemShut {NoStop}%
\bibitem [{\citenamefont {Ba{\~n}os}\ \emph {et~al.}(2012)\citenamefont
  {Ba{\~n}os}, \citenamefont {Cruz}, \citenamefont {Fernandez}, \citenamefont
  {Gil-Narvion}, \citenamefont {Gordillo-Guerrero}, \citenamefont {Guidetti},
  \citenamefont {I{\~n}iguez}, \citenamefont {Maiorano}, \citenamefont
  {Marinari}, \citenamefont {Martin-Mayor} \emph {et~al.}}]{JanusPNAS}%
  \BibitemOpen
  \bibfield  {author} {\bibinfo {author} {\bibfnamefont {R.~A.}\ \bibnamefont
  {Ba{\~n}os}}, \bibinfo {author} {\bibfnamefont {A.}~\bibnamefont {Cruz}},
  \bibinfo {author} {\bibfnamefont {L.~A.}\ \bibnamefont {Fernandez}}, \bibinfo
  {author} {\bibfnamefont {J.~M.}\ \bibnamefont {Gil-Narvion}}, \bibinfo
  {author} {\bibfnamefont {A.}~\bibnamefont {Gordillo-Guerrero}}, \bibinfo
  {author} {\bibfnamefont {M.}~\bibnamefont {Guidetti}}, \bibinfo {author}
  {\bibfnamefont {D.}~\bibnamefont {I{\~n}iguez}}, \bibinfo {author}
  {\bibfnamefont {A.}~\bibnamefont {Maiorano}}, \bibinfo {author}
  {\bibfnamefont {E.}~\bibnamefont {Marinari}}, \bibinfo {author}
  {\bibfnamefont {V.}~\bibnamefont {Martin-Mayor}},  \emph {et~al.},\
  }\href@noop {} {\bibfield  {journal} {\bibinfo  {journal} {Proceedings of the
  National Academy of Sciences}\ }\textbf {\bibinfo {volume} {109}},\ \bibinfo
  {pages} {6452} (\bibinfo {year} {2012})}\BibitemShut {NoStop}%
\bibitem [{\citenamefont {Leuzzi}\ \emph {et~al.}(2009)\citenamefont {Leuzzi},
  \citenamefont {Parisi}, \citenamefont {Ricci-Tersenghi},\ and\ \citenamefont
  {Ruiz-Lorenzo}}]{LPRR09}%
  \BibitemOpen
  \bibfield  {author} {\bibinfo {author} {\bibfnamefont {L.}~\bibnamefont
  {Leuzzi}}, \bibinfo {author} {\bibfnamefont {G.}~\bibnamefont {Parisi}},
  \bibinfo {author} {\bibfnamefont {F.}~\bibnamefont {Ricci-Tersenghi}}, \ and\
  \bibinfo {author} {\bibfnamefont {J.}~\bibnamefont {Ruiz-Lorenzo}},\
  }\href@noop {} {\bibfield  {journal} {\bibinfo  {journal} {Physical review
  letters}\ }\textbf {\bibinfo {volume} {103}},\ \bibinfo {pages} {267201}
  (\bibinfo {year} {2009})}\BibitemShut {NoStop}%
\bibitem [{\citenamefont {Angelini}\ and\ \citenamefont {Biroli}(2014)}]{AB14}%
  \BibitemOpen
  \bibfield  {author} {\bibinfo {author} {\bibfnamefont {M.~C.}\ \bibnamefont
  {Angelini}}\ and\ \bibinfo {author} {\bibfnamefont {G.}~\bibnamefont
  {Biroli}},\ }\href@noop {} {\bibfield  {journal} {\bibinfo  {journal} {arXiv
  preprint arXiv:1409.1011}\ } (\bibinfo {year} {2014})}\BibitemShut {NoStop}%
\bibitem [{foo({\natexlab{e}})}]{footnoteG}%
  \BibitemOpen
  \href@noop {} {\emph {\bibinfo {title} {\emph{It could be interesting to see
  if it is possible to find a perturbative regime analogous to what has been
  found in \cite{PT12}}}}}\BibitemShut {NoStop}%
\bibitem [{\citenamefont {Franz}(2005)}]{Fr05}%
  \BibitemOpen
  \bibfield  {author} {\bibinfo {author} {\bibfnamefont {S.}~\bibnamefont
  {Franz}},\ }\href@noop {} {\bibfield  {journal} {\bibinfo  {journal} {Journal
  of Statistical Mechanics: Theory and Experiment}\ }\textbf {\bibinfo {volume}
  {2005}},\ \bibinfo {pages} {P04001} (\bibinfo {year} {2005})}\BibitemShut
  {NoStop}%
\bibitem [{\citenamefont {Parisi}\ and\ \citenamefont
  {Temesv{\'a}ri}(2012)}]{PT12}%
  \BibitemOpen
  \bibfield  {author} {\bibinfo {author} {\bibfnamefont {G.}~\bibnamefont
  {Parisi}}\ and\ \bibinfo {author} {\bibfnamefont {T.}~\bibnamefont
  {Temesv{\'a}ri}},\ }\href@noop {} {\bibfield  {journal} {\bibinfo  {journal}
  {Nuclear Physics B}\ }\textbf {\bibinfo {volume} {858}},\ \bibinfo {pages}
  {293} (\bibinfo {year} {2012})}\BibitemShut {NoStop}%
\end{thebibliography}
%

\end{document}